\documentclass{llncs}

\usepackage{graphicx}
\usepackage{algorithm,algorithmic}
\usepackage{slashbox}
\usepackage{float}
\usepackage{amssymb}
\usepackage{amsmath}
\usepackage{amsfonts}
\usepackage{wasysym}
\usepackage{color} 
\usepackage{soul}
 \usepackage{url}
 \usepackage{breakurl} 
\begin{document}

\title{Energy Efficient Scheduling of Application Components via Brownout and Approximate Markov Decision Process}

%\author{\IEEEauthorblockN{Minxian Xu and Rajkumar Buyya}
%	\IEEEauthorblockA{\textbf{Clou}d Computing and \textbf{D}istributed \textbf{S}ystems (CLOUDS) Lab\\
%		School of Computing and Information Systems\\
%		The University of Melbourne, Australia\\
%		Email: minxianx@student.unimelb.edu.au, rbuyya@unimelb.edu.au}}

\author{Minxian Xu and Rajkumar Buyya}

\institute{\textbf{Clou}d Computing and \textbf{D}istributed \textbf{S}ystems (CLOUDS) Laboratory\\
	School of Computing and Information Systems\\
	The University of Melbourne, Australia \\
	\textbf{minxianx@student.unimelb.edu.au, rbuyya@unimelb.edu.au}
}

%\author{Minxian~Xu~\IEEEmembership{}
%	and~Rajkumar Buyya,~\IEEEmembership{Fellow,~IEEE}% <-this % stops a space
%	\IEEEcompsocitemizethanks{\IEEEcompsocthanksitem M. Xu, A. V. Dastjerdi and R. Buyya are with Cloud Computing and Distributed Systems (CLOUDS) lab, Department of Computing and Information Systems,
%		University of Melbourne, Australia, 3010
%		.\protect\\
%		% note need leading \protect in front of \\ to get a newline within \thanks as
%		% \\ is fragile and will error, could use \hfil\break instead.
%		E-mail: xianecisp@gmail.com
%	}}% <-this % stops a space
\maketitle

\begin{abstract}

 Unexpected loads in Cloud data centers may trigger overloaded situation and performance degradation. To guarantee system performance, cloud computing environment is required to have the ability to handle overloads. The existing approaches, like Dynamic Voltage Frequency Scaling and VM consolidation, are effective in handling partial overloads, however, they cannot function when the whole data center is overloaded. Brownout has been proved to be a promising approach to relieve the overloads through deactivating application non-mandatory components or microservices temporarily. 
\color{black} Moreover, brownout has been applied to reduce data center energy consumption.  It shows that there are trade-offs between energy saving and discount offered to users (revenue loss) when one or more services are not provided temporarily.
In this paper, we propose a brownout-based approximate Markov Decision Process approach to improve the aforementioned trade-offs. The results based on real trace demonstrate that our approach saves 20\% energy consumption than VM consolidation approach. Compared with existing energy-efficient brownout approach, our approach reduces the discount amount given to users while saving similar energy consumption. 
\color{black}

\end{abstract}

\begin{keywords}
	%Cloud Data Centers, 
	Cloud Energy Efficiency, Application Component, Microservices, Brownout, Markov Decision Process
\end{keywords}

%\IEEEpeerreviewmaketitle

\section{Introduction}\label{sec:Introduction}

% Cloud computing provides compelling features such as pay-as-you-go pricing model, low operation cost, high scalability and easy access, and it is viewed as a new paradigm in IT industry \cite{Buyya}. 
%This makes Cloud computing attractive to business owners as it eliminated the requirement for users to plan ahead for provisioning, and allows enterprises to start with minimum and request resources on demand. Providers like Amazon, Microsoft, IBM and Google have established data centers to support cloud applications around the world, and aimed to ensure that their services are flexible and suitable for the needs of the end-users.
%However, 
\color{black}Given the scenario that budget and resource are limited, overloaded situation may lead to performance degradation and resource saturation, in which some requests cannot be allocated by providers. 
Thus, some users may experience high latencies, and others may even not receive services at all \cite{Klein}, which directly affects the requests that have Quality of Service (QoS) constraints.
Unfortunately, current resource management approaches, like Dynamic Voltage Frequency Scaling (DVFS) \cite{Kim} and VM consolidation \cite{Pecero}, cannot function when the \textbf{holistic} data center is overloaded.  
\color{black}The saturated resource not only brings over-utilized situation to hosts, but also causes high energy consumption. 

Energy consumed by the cloud data centers has currently become one of the major concerns of the computing industry. 
%The growth and development of complex data-driven applications have promulgated the creation of huge data centers, which heightens the energy consumption \cite{Kaur}. 
It is reported that U.S. data centers will consume 140 billion kWh of electricity annually by 2020, which equals to the annual output of about 50 brown power plants \cite{Pierre}.
Analysts also forecast that data centers will roughly triple the amount of electricity consumed in the next decade \cite{Tom}.
The servers hosted in data centers dissipate heat and need to be maintained in a fully air-conditioned and engineered environment. 
Though the cooling system is already efficient, servers remain one of the major energy consumers. %Therefore, reducing server energy consumption has become a main concern of researchers. 
One of the main reasons of high energy consumption lies in that computing resource are not efficiently utilized by server applications. Currently, building applications with microservices provides a more efficient approach to utilize infrastructure resource.

\color{black}Applications can be constructed via set of self-contained components which are also called microservices. The components encapsulate its logic and expose its functionality through interfaces, which makes them flexible to be deployed and replaced. With components or microservices, developers and users can benefit from their technological heterogeneity, resilience, scalability, ease of deployment, organizational alignment, composability and optimization for replaceability \cite{Newman}. This also brings the advantage of more fine-grained control over the application resource consumption.

%Therefore, in the brownout approach, it is feasible to downgrade user experience by disabling part of non-mandatory application components to relieve the over-utilized condition and reduce energy consumption. 

Therefore, we take advantage of a paradigm called \textbf{brownout} \cite{Klein} to handle with overloaded situation and save energy. It is inspired by the concept of brownout in electric grids and originates from the voltage shutdown that copes with emergency cases, in which light bulbs emit fewer lights and consume less power \cite{Durango}. In Cloud scenario, brownout can be applied to applications components or microservices that are allowed to be disabled temporarily. %in which fewer requests with QoS constraints are affected by overloads.

It is common that application components or microservices have this brownout feature. A brownout example for online shopping system is introduced in \cite{Klein}, in which the online shopping application provides a recommendation engine to recommend products that users may be interested in. The recommendation engine component helps service provider to increase profits, but it is not required to be running all the time. Recommendation engine also requires more resource in comparison with other components. Accordingly, with brownout, under overloaded situation, the recommendation engine could be deactivated to serve more clients who require essential services and have QoS constraints. \color{black} Another example is the online document process application that contains the components for spell checking and  report generation. These components are not essential to run all the time and can be deactivated for a while to reduce resource utilization.  
Apart from these two examples, brownout is available for other application components or microservices that are not required to be available all the time. 
%What the service providers need to spend efforts on are identifying the optional components and determining their discount when  deactivated. 
 %Our motivation is to investigate the trade-off between energy consumption and discount, as well as offering different component selection policies.
\color{black}

In this paper, we consider component-level control in our system model. The model could also be applied to container or microservices architecture. We model the application components as \color{black}either mandatory or optional\color{black}, and if required, optional components can be deactivated. By deactivating the optional components selectively and dynamically, the application utilization is reduced to save total energy consumption. While under market scenario, service provider may provide discount for users as one or more services are deactivated. 

\color{black}In our scenario, the meaning of discount is not limited to the discount offered to users. Additionally, it can also be modelled as the revenue loss of service providers (i.e. SaaS service providers) that they charge lower price for services under brownout. For
example, in an online shopping system, the recommendation engine helps the service provider to improve their revenue by recommending similar products. If the recommendation engine is deactivated, the service provider is unable to obtain
the revenue from recommendation engine.\color{black}
%The discount may also be the revenue loss of service provider. For instance, if the recommendation engine is deactivated, service provider loses the revenue increased by the recommendation engine. In this paper, we also note the revenue loss as discount.

The key \textbf{contributions} of this paper are: \color{black}our approach considers the trade-offs between saved energy and the discount that is given to a user if components or microservices are deactivated; \color{black}we propose an efficient algorithm based on brownout and approximate Markov Decision Process that considers the aforementioned trade-offs and achieves better trade-offs than baselines.

\color{black}The remainder of this paper is organized as follows: after discussing the related work in Section 2, we present the brownout system model and problem statement in Section 3. Section 4 introduces our proposed brownout-based Markov Decision Process approach, and Section 5 demonstrates the experimental results of our proposed approach. The summary along with the future work are concluded in Section 6. \color{black}

\section{Related Work}

%It is an essential requirement for Cloud providers to reduce energy consumption, as it can both decrease operating costs and improve system reliability. Data centers can consume from 10 to 100 times more power per square foot than typical office building. 
A large body of literature has focused on reducing energy consumption in cloud data centers, and the dominant categories for solving this problem are VM consolidation and Dynamic Voltage Frequency Scaling (DVFS). 
%Server virtualization and VM consolidation have been applied in approaches  \cite{Mastroianni} \cite{Corradi} \cite{Salehi}.  DVFS techniques have been adopted to reduce energy consumption in \cite{Laszewski} \cite{Kim} \cite{Hanumaiah}.

VM consolidation is viewed as an act of combining into an integral whole, which saves energy by allocating work among fewer machines and turning off unused machines \cite{Pecero}. Using this approach, VMs allocated to underutilized hosts are consolidated to other servers and the remaining hosts are transformed into low power mode. Mastroianni et al. \cite{Mastroianni} presented a self-organizing and adaptive approach for consolidation of VMs CPU and RAM resource, which is driven by probabilistic processes and local information. 
Corradi et al. \cite{Corradi} considered VM consolidation in a more practical viewpoint related to power, CPU and networking resource sharing and tested VM consolidation in OpenStack, which shows VM consolidation is a feasible solution to lessen energy consumption. 
%\color{black}Salehi et al. \cite{Salehi} proposed a VM consolidation based approach, which is an adaptive energy management policy that preempts VMs to reduce the energy consumption according to user-specific performance constraints and used fuzzy logic for obtaining appropriate decisions. \color{black}
%Ferdaus et al. \cite{Ferdaus} presented a VM consolidation algorithm based on Ant Colony Optimization to reduce both energy consumption and resource wastage.

The DVFS technique introduces a trade-off between computing performance and  energy consumed by the server. The DVFS technique lowers the frequency and voltage when the processor is lightly loaded, and utilizes maximum frequency and voltage when the processor is heavily loaded. 
%\color{black}Von et al. \cite{Laszewski} introduced a power-aware scheduling algorithm based on DVFS-enabled cluster. \color{black}
Kim et al. \cite{Kim} proposed several power-aware VM schemes based on DVFS for real-time services.  Hanumaiah et al. \cite{Hanumaiah} introduced a solution that considers DVFS, thread migration and active cooling to control the cores to maximize overall energy efficiency. 
%Catalan et al. \cite{catalan} conducted experiments on IBM Power8 to evaluate the energy consumption effects caused by Voltage Frequency Scaling for matrix vector products.

%Brownout was originally applied to prevent blackouts through voltage drops in case of emergency. 
Most of the proposed brownout approaches in Cloud scenarios focused on handling overloads or overbooking rather than energy efficiency perspective.
Klein et al. \cite{Klein} firstly borrowed the approach of brownout and applied it to cloud applications, aiming to design more robust applications under unpredictable loads. Tomas et al. \cite{Tomas} used brownout along with overbooking to ensure graceful degradation during load spikes and avoid overload. 
%Durango et al. \cite{Durango} introduced novel load balancing strategies for applications by supporting brownout. 
In a brownout-compliant application or service, the optional parts are identified by developers, and a control knob called \textbf{dimmer} that controls these optional parts is also introduced. The dimmer value represents a certain probability given by a control variable and shows how often these optional parts are executed. Moreover, a brownout controller is also required to adjust the dimmer value.  

\color{black}Markov Decision Process (MDP) is a discrete time stochastic optimization approach and provides a way to solve the multiple state probabilistic decision-making problem, which has been adopted to solve resource management problems in Cloud scenarios. \color{black}Toosi et al. \cite{Adel} used finite MDP for requests admission control in Clouds, while their objective is maximizing revenues rather than reducing power consumption. Han et al. \cite{Han} applied MDP to determine VM migration for minimizing energy consumption, while our work is adopting MDP to determine the deactivation of application components. 

In our previous work \cite{Xu}, several heuristic policies were proposed to find the components that should be deactivated and investigated the trade-offs between energy and discount. In this paper, we adopt approximate
 MDP to improve the aforementioned trade-offs.

%Software Consolidation as an Efficient Energy
%and Cost Saving Solution for a SaaS/PaaS
%Cloud Model Alain Tchana1

%The Impact of Voltage-Frequency Scaling for the
%Matrix-Vector Product on the IBM POWER8
%Sandra Catal´an1

\section{System Model and Problem Definition}

\subsection{System Model}
  Our system model is presented in Fig. 1 and it consists of the following entities:  
  
  \textbf{Users}:  Users submit service requests to cloud data centers. The users entity contains user information and requested applications (services). 
  
  \textbf{Applications}: The applications provide different services for users and are consisted of a set of components, which are identified as mandatory or optional. 
  
  \textbf{Mandatory component}: The mandatory component keeps running all the time when the application is launched. 
  %is always running (activated) when the application is executed. 
  
  \textbf{Optional component}: The optional component can be set as activated or deactivated according to the system status. These components have parameters like utilization $u(App_c)$ and discount $d(App_c)$. Utilization indicates the amount of  utilization, and discount represents the amount of discount that is offered to the users (or revenue loss of service provider). The operations of optional components are controlled by the \textbf{brownout controller}, which makes decisions based on the system overloaded status and brownout algorithm. 
  
  \color{black}To adapt the dimmer to our model, different from the dimmer in \cite{Klein} that requires a dimmer per application, our dimmer  is only applied to the applications with optional components. Rather than response time, another adaptation is that our dimmer value is computed based on the number of overloaded hosts and adapts to the severity of overloaded events (more details are presented in Section 4.1)\color{black}.
  
  %The components can also be \textit{connected}, which means that they  communicate with each other and there are data dependencies between them. Therefore, we consider that if a component is deactivated, then all its connected optional components would also be set as deactivated. For example in Fig. 1, if Com3 in Application \#1 is deactivated, Com2 should also be deactivated; in Application \#2, if Com1 is deactivated, Com3 should also be deactivated; in Application \#n, if Com4 is deactivated, Com3 should also be deactivated, but Com2 is still working (Com1 is connected with Com3, but Com1 is mandatory, so it is not deactivated).
  
  \textbf{Cloud Providers}: Cloud providers offer physical resources to meet service demands, which host a set of VMs or containers to run applications.

 \begin{figure}[t]
 	\centering
 	\includegraphics[width=0.7\linewidth]{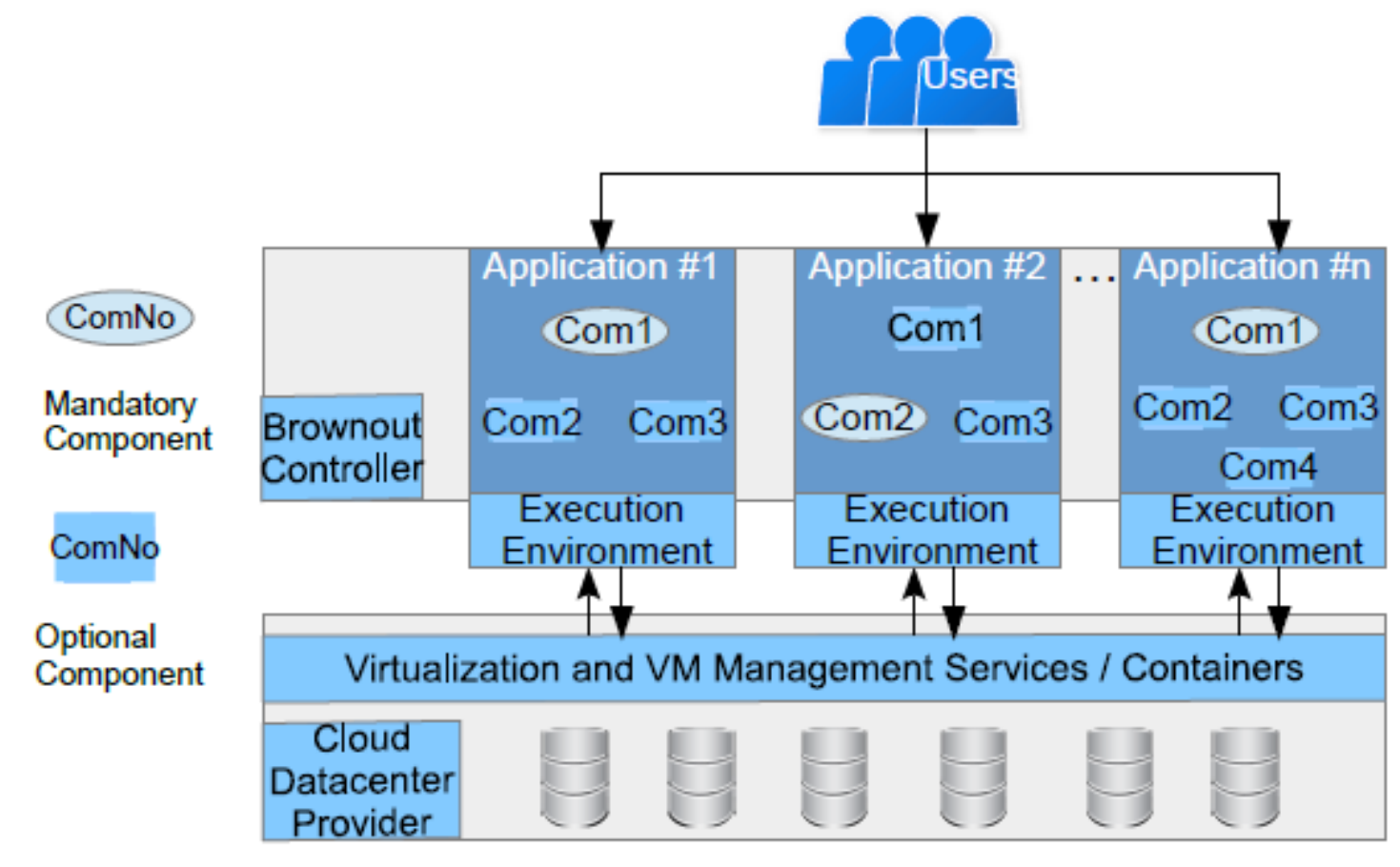}
 	\caption[Brownout Architecure]{System Model with Brownout}
 	
 	\label{fig:BrownoutArchi}
 \end{figure}

\subsection{Power Model}

We adopt the servers power model derived from \cite{Zheng}. The power of server $i$ is $P_i(t)$ that is dominated by the CPU utilization:
\begin{equation}
	P_i(t) = 	\begin{cases}
		P_i^{idle} + \sum_{j=1}^{N_i} u(VM_{i,j}(t)) \times P_i^{dynamic}   & ,N_i > 0\\
		0 & ,N_i = 0
	\end{cases}
\end{equation}

$P_i(t)$ is composed of idle power and dynamic power. The idle power is regarded as constant and the dynamic power is linear to the total CPU utilization of all the VMs on the server \cite{Zheng}. If no VM is hosted on a server,  the server is turned off to save power. $VM_{i,j}$ refers to the $j$th VM on server $i$, $N_i$ means the number of VMs assigned to server $i$. And $u(VM_{i,j}(t))$ refers to the VM utilization at time interval $t$, which is represented as:

%Total energy consumption under policy $\Omega$:

%\begin{equation}
%E_{\Omega} \triangleq lim_{T\rightarrow + \infty} \sum_{t=1}^{T}E_{total}(t)
%\end{equation}

\begin{equation}
u(VM_{i,j}(t)) = \sum_{c=1}^{C_j}u(App_c)
\end{equation}
where $C_j$ is the number of application components on VM, and $u(App_c)$ is the utilization of application component $c$ when it is activated.

Then the total energy consumption during time interval $t$, with $M$ servers is:

\begin{equation}
E(t) = \sum_{i=1}^{M}\int_{t-1}^{t} P_{i}(t) dt
\end{equation}

%\color{black} \textbf{Notes:} In our power model, we have several assumptions related to the energy consumption by turning on/off hosts and deactivating/activating application components: 1) We assume that the time required to turn off/on hosts is less than a scheduling time interval (like 5 minutes). With this assumption, when the machine remains idle in a time interval, it will be turned off. When the host is turned off/on, the host is assumed to be running with the idle power. 2) Our application components can be implemented as containers or microservices, which has low overheads to start or stop components. Therefore, in our model, we assume that the utilization and time delay of deactivating and activating components is negligible.
\color{black} \textbf{Notes:} In our power model, we assume that the time required to turn on/off hosts (including the time to deactivate and activate components) is lees than a scheduling time interval (like 5 minutes). When the host is turned off/on, the host is assumed to be consuming the idle power.  

\color{black}

\subsection{Discount Amount}
\color{black}%As mentioned in Section I, the discount amount represents the penalty given back to the users or the revenue loss of service provider when services are temporarily deactivated. In this paper, we note them as discount.
As introduced in Section 1, the meaning of discount could be either the discount offered to users or the revenue loss of service providers that they charge lower price for services under brownout. 
%For example, in the online shopping system, the recommendation engine helps the service provider to improve their revenue by recommending similar products. If the recommendation engine is deactivated, the service provider is unable to obtainthe revenue from recommendation engine, which is revenue loss or another representation of discount. 
In this paper, we note them as discount.
\color{black}

%Total discount amount under policy $\Omega$, with $n$ applications:
%\begin{equation}
%D_{\Omega} \triangleq lim_{T\rightarrow + \infty} \sum_{t=1}^{T}D_{total}(t)
%\end{equation}
The total discount amount at time interval $t$ is modeled as the sum of discount of all deactivated application components at $t$:

 \begin{equation}
 D(t) = \sum_{i=1}^{M} \sum_{j=1}^{N_i}d(VM_{i,j}(t))
 \end{equation} 
 
 where $D(t)$ is the total discount amount at $t$ that obtained from all VMs on hosts, $N_i$ is the number of VMs assigned to server $i$, $M$ is the number of servers. The individual discount $d(VM_{i,j}(t))$  is the sum of discount amount of deactivated application components $d(App_c)$ of $VM_{i,j}$, which is shown in Equation (5):

 \begin{equation}
 d(VM_{i,j}) = \sum_{c=1}^{C_j} d(App_c)
 \end{equation} 
where $C_j$ is the number of application components hosted on $VM_j$, and only the deactivated components are counted. 
%$d(VM_{i,j})$  is the discount offered from $VM_{i,j}$ on server $i$.
% and $D_i$ is the total discount amount on server $i$. 

\subsection{Problem Definition}

Let $Q(t) \in \mathbb{Q}$, where $\mathbb{Q} = {\eta_1,\ldots, \eta_{\mathbb{|Q|}}}$, $\eta_i \in \mathbb{Q}$. The $Q(t)$ is a combination of two vectors: energy consumption vector $E(t)$ and discount amount vector $D(t)$\color{black}, representing the possible energy consumption and discount amount at different system states\color{black}. Let $C(t)$ to be all the application component states at $t$, we have

\noindent\textbf{Definition 1}	\textit{The system state at time interval $t$ can be specified as:} 
	\begin{equation}
	S(t) \triangleq [Q(t), C(t)]
	\end{equation}
\color{black}The system state $S(t)$ contains the energy consumption and discount amount as well as their corresponding application components states \color{black}.

At each time interval, we calculate the state information as:

\begin{equation}
g(t) = E(t) + \lambda D(t)
\end{equation}
where $\lambda$ is the weight of discount. \color{black}The higher $\lambda$ implicates that more weights are given to the discount amount\color{black}.
In the whole scheduling period $T$ under policy $\pi$, our optimization objective is:
\begin{equation}
	\min_{\pi} \qquad g(\pi) = \sum_{t=0}^{T}[E(t) + \lambda D(t)] 
\end{equation}

\section{Proposed Approach}

\subsection{Approximate Markov  Decision Process Model}
 To adopt the Markov model, we assume that the workload satisfies the Markov property, which means the state transitions caused by workloads are memoryless. Our experiments are conducted with Planetlab workload, which has been validated to satisfy Markov chain property \cite{Beloglazov}. In our model, we assume that the probability of application components to transfer their states at the next time period only depends on the workloads of the current time period and independent on earlier states. We formulate our problem as finite horizon MDP that we investigate a fixed length of time.

Then we can solve our objective function by using Bellman equation \cite{Bellman}:
\begin{equation}
V^*(S_i) = \arg\min_{\gamma \in \mathbb{R}}[{g(S_i) + \sum_{S_j \in S} Pr [S_j | S_i, \gamma] V^*(S_j)}]
\end{equation}
$g(S_i)$ is the instant cost under system state $S_i$, and
$V^*(S_i)$ is the expected energy consumption and discount obtained from $S_j$ to $S_i$.
We also denote $\gamma(t) \triangleq [\gamma_1(t), \ldots, \gamma_n(t)] \in \mathbb{R}$ as the operations (activation or deactivation actions) for application components.
$V^*(S_i)$ can be found by iteratively obtaining minimum energy consumption and discount until convergence.

Let $\hat{p_{i,j}}$ denote the estimated transition probability that the application component changes its state. The transition probability is computed as:

%\begin{equation}
%\begin{aligned}
%Pr[S(t+1)|S(t), \pi(S(t))] & = Pr[Q(t+1), C(t+1)| Q(t), C(t), \pi(S(t))]  \\
%& = Pr[Q(t+1)|Q(t)]Pr[C(t+1)|C(t), \pi(S(t))] \\
%& = Pr[Q(t+1) = \eta_i| Q(t) = \eta_j] \\
%& = p_{i,j} 
%\end{aligned}
%\end{equation}

%$A(t)$ is application components states (deactivated or activated).

\begin{equation}
\hat{p}_{i,j} = \sqrt{\frac{\hat{M}}{M}} \times Pr(\frac{u(App_c)}{d(App_c)} = z_{C})
\end{equation}
$Pr(\frac{u(App_c)}{d(App_c)} = z_{C})$ is the probability that the ratio of component utilization and discount  $\frac{u(App_c)}{d(App_c)}$ falls into category $z_{C}$\color{black}.  We divide the probability into $C$ (the maximum number of components on a VM) categories. For all the components with the probability falls into the same category, we apply the same operation. To avoid the curse of dimension, noted by \cite{Han}, we adopt key states to reduce state space. With key states, the component states on a VM is reduced to the maximum number of components on a VM as $|C|$\color{black}. 
 $\hat{M}$ is the estimated number of overloaded hosts, which is calculated based on a sliding window \cite{AntonTPDS}. \color{black} The advantage of sliding window is to give more weights to the values of recent time intervals\color{black}. Let $L_w$ to be the window size, and $N(t)$ to be the number of overloaded hosts at $t$, we estimate $\hat{M}$ as:

\begin{equation}
\hat{M}(L_w) = \frac{1}{L_w}\sum_{t=0}^{L_w-1} N(t)
\end{equation}
%where $L_w$ is the window size.

%The error rate $WER_{i,j}$ of estimated value is related to window size: 
%\begin{equation}
%ER_{i, j} = \frac{1}{T}\sum_{t=1}^{T}[\hat{p}_{i,j}(t) - p_{i,j}(t)]^2
%\end{equation}
%which shows accuracy can be improved with larger window size.

 We denote the states as key states $S_k$ as described above. With proof in \cite{Han}$, \forall S_i \in S_k$ for all the VMs, the equivalent Bellman's equation in Equation (9) can be approximately formulated as:
\begin{equation}
%\footnotesize 
\small
V^*(S_i) \approx \sum_{m=1}^{M}\sum_{n=1}^{N_m}(g(S_i) + \arg\min_{\gamma_n \in \mathbb{R}_n} \{\sum_{S_j \in S_k}Pr[S_j|S_i, \gamma_n]\widetilde{V^*_n}(S_j)\})
\end{equation}
The state spaces thus are reduced to polynomial with linear approximation. The $M$ is the number of hosts and $N_m$ is the number of VM assigned to server $m$. 

\subsection{Brownout Algorithm based on Markov Decision Process (BMDP)}
Our novel brownout algorithm is embedded within a VM placement and consolidation algorithm. We adopt the VM placement and consolidation algorithm (PCO) proposed in \cite{Beloglazov}, which is also one of our baselines in Section 5.

The PCO algorithm is a heuristic to reduce energy consumption through VM consolidation. In the initial VM placement phase, PCO sorts all the VMs in decreasing order by their current CPU utilization and allocates each VM to the host that increases the least power consumption due to this allocation. In the VM consolidation phase, PCO optimizes VM placement by separately picking VMs from over-utilized and under-utilized hosts to migrate, and finding new placements for them. 
After migration, the over-utilized hosts are not overloaded any more and the under-utilized hosts are switched to sleep mode. 

Our brownout algorithm based on approximate Markov Decision Process is shown in Algorithm 1 and includes 6 steps:

\textbf{1) System initialization (lines 1-2):}
Initializing the system configurations, including overloaded threshold $TP$, dimmer value $\theta_t$, vector $\mathbb{Q}$ that contains the $D(t)$ and $E(t)$ information, as well as objective states $\mathbb{S}_d$, and applying VM placement algorithm in PCO to initialize VM placement.

\textbf{2) Estimating transition probability of each application component (lines 3-14):} At each time interval, the algorithm firstly estimates the number of overloaded hosts. \color{black} The dimmer value is computed as $\sqrt{\frac{\hat{M}}{M}}$, which is adaptive to the number of overloaded hosts. If no host is overloaded, the value is 0 and no component is deactivated. If there are overloaded hosts, the  transition probabilities of application components are computed using Equation (10)\color{black}. 

\textbf{3) Finding the states that minimize the objective function (lines 15-17):} Traversing all the key states by value iteration according to \color{black}Equation (12)\color{black}, where $D^{'}(t)$ and $E^{'}(t)$ are the temporary values at the current state.

\textbf{4) Updating system information (lines 18-20):} The algorithm updates the obtained energy consumption and discount values if $g(t)$ in Equation (7) is reduced, and records the optimized states. The current states are substituted by the state with lower $g(t)$. 

\textbf{5) Deactivating the selected components (line 22):} The brownout controller deactivates the selected components to achieve objective states.

\textbf{6) Optimize VMs placement (line 24)：} The algorithm uses the VM consolidation approach in PCO to optimize VM placement via VM consolidations.

 \begin{algorithm}[t]
 	%\large
 	%\normalsize
    \scriptsize
 	\caption{Brownout based Markov Decision Process Algorithm (BMDP)}
 	\begin{algorithmic}[1]
 		\renewcommand{\algorithmicrequire}{\textbf{Input:}}
 		\renewcommand{\algorithmicensure}{\textbf{Output:} }
 		\REQUIRE  host list $hl$ with size $M$, VM list, application components information, overloaded power threshold $TP$, dimmer value $\theta_t$ at time $t$, destination states $S_d(t)$, energy consumption $E(t)$ and discount amount $D(t)$ in $\mathbb{Q}$
 		%deactivated component list $dcl_{i, j, t}$ of $VM_{j}$ on host $h_i$, power model of host $HPM$, VM utilization model $VUM$, component selection policy $CSP$
 		\ENSURE  total energy consumption, discount amount
 		\STATE $TP$ $\leftarrow$ 0.8;  $\theta_t$ $\leftarrow$ 0; $\forall E(t), \forall D(t) \in \mathbb{Q} \leftarrow max$; $S_d(t) \in \mathbb{S}_d \leftarrow NULL$
% 		\STATE
% 		\STATE   
% 		\STATE　
 		\STATE use PCO algorithm to initialize VMs placement
 %		\STATE $dcl_{i, j, t}$ $\leftarrow$ NULL
 		
 		\WHILE {true}
 		
 		\FOR{$t \leftarrow 0$ to $T$}
 		
% 		\IF{$n_t > 0$}
 		\STATE  $\theta_t$ $\leftarrow$ = $ \sqrt{\frac{\hat{M}_t}{M}}$

 		\FORALL {$h_i$ \textbf{in} $hl$}
 		\IF{$h_i$ is overloaded}
 		\FORALL {$VM_{i, j}$ on $h_i$}
 		\FORALL {$App_c$ on $VM_{i,j}$}
 		\STATE $Pr(App_c) \leftarrow \theta_t \times Pr(\frac{u(App_c)}{d(App_c)}=z_{C})$ 
 		\ENDFOR
 		\ENDFOR
 		
% 		\STATE $P^{r}_i$ $\leftarrow$ $\theta_t$ $\times$ $P_{h_i}$
% 		\STATE $u_{h_i}^{r}$ $\leftarrow$ $HPM$($P^{reduced}_i$)
% 		\STATE $u_{VM_{i, j}}^{r}$ $\leftarrow$ $VUM$($u_{h_i}^{reduced}$, $vl$)
% 		\STATE $dcl_{i, j, t}$ $\leftarrow$ $CSP$($u_{VM_{i, j}}^{reduced}$)
% 		\STATE $D_i \leftarrow D_i + d(VM_{i, j})$
 		
 		\ENDIF
 		\ENDFOR
		 
		\FORALL {$S_j(t) \in S_k(t)$}
		\STATE $V^*(S_i) = \sum_{m = 1}^{m = M}\sum_{n=1}^{n=N_m}(g(S_i) + \min_{\gamma_n \in \mathbb{R}_n} \{\sum_{S_j \in S_k}Pr[S_j|S_i, \gamma_n]\widetilde{V^*_n}(S_j)\})$
		\STATE $g(t) = E^{'}(t) + \lambda D^{'}(t)$ 
		\IF{$g(t) < E(t) + \lambda D(t)$}  
		\STATE $E(t) \leftarrow E^{'}(t)$ ; $D(t) \leftarrow D^{'}(t)$ ; $S_d(t) \leftarrow S_j(t)$ 
%		\STATE 
%		\STATE 
		\ENDIF
		\ENDFOR

		\STATE deactivate the selected components to achieve state $S_d(t)$
% 		\ELSE 
% 		\STATE activate deactivated components
% 		\ENDIF
 		 \ENDFOR
 		
 		\STATE use VM consolidation in PCO algorithm to optimize VM placement
 		\ENDWHILE

 	\end{algorithmic} 
 \end{algorithm}
 
The complexity of the BMDP algorithm at each time interval is consisted of the brownout part and VM consolidation part. The complexity of the transition probability computation is $O(C\cdot N \cdot M) $, where $C$ is the maximum number of components in all applications, $N$ is the maximum number of VMs on all the hosts and $M$ is the number of hosts.
With the key states, the space state of the MDP in brownout part is $O(C\cdot N \cdot M) $.  According to Equation (12), the actions are reduced to $O(C \cdot N \cdot M)$, so the overall MDP complexity is $O(C^2\cdot N^2 \cdot M^2)$.	
The complexity of the PCO part is $O(2M) $ as analyzed in \cite{Beloglazov}. 
Therefore, the overall complexity is $O(C\cdot M\cdot N + C^2\cdot N^2 \cdot M^2 + 2M)$ or equally $O(C^2\cdot N^2 \cdot M^2)$.

\section{Performance Evaluation}
\color{black}\subsection{Methodology}\color{black}
We use the CloudSim framework \cite{Buyya2} to simulate a cloud data center. The data center contains two types of hosts and four types of VMs that are modeled based on current offerings in EC2 as shown in Table 1. The power models of the adopted hosts are derived from IBM System x3550 M3 with CPU Intel Xeon X5670 and X5675 \cite{SPEC}
. \color{black}We set the time required to turn on/off hosts as 0.5 minute\color{black}.
%\color{black}, and their power consumption at different utilization levels are shown in Table 1. For example, in the power model with Intel Xeon X5670 CPU, if the utilization is decreased from 100\% to 80\%, then the power is reduced from 247 to 211 Watts.

%\begin{table}[t]
%	\color{black}
%	
%	\centering
%	\caption{Power consumption of servers in Watts}
%	
%	\label{my-label}
%	\resizebox{0.85\textwidth}{!}{%
%		\begin{tabular}{|c|c|c|c|c|c|c|c|c|c|c|c|}
%			\hline
%			\textbf{Servers} & \textbf{\begin{tabular}[c]{@{}c@{}}0\% \\ (sleep mode)\end{tabular}} & \textbf{10\%} & \textbf{20\%} & \textbf{30\%} & \textbf{40\%} & \textbf{\begin{tabular}[c]{@{}c@{}}50\% \\ (idle)\end{tabular}} & \textbf{60\%} & \textbf{70\%} & \textbf{80\%} & \textbf{90\%} & \textbf{\begin{tabular}[c]{@{}c@{}}100\% \\ (max)\end{tabular}} \\ \hline
%			\begin{tabular}[c]{@{}c@{}}IBM x3550 M3 \\ (Interl Xeon X5670 CPU)\end{tabular} & 66 & 107 & 120 & 131 & 143 & 156 & 173 & 191 & 211 & 229 & 247 \\ \hline
%			\begin{tabular}[c]{@{}c@{}}IBM x3550 M3 \\ (Intel Xeon X5675 CPU)\end{tabular} & 58.4 & 98 & 109 & 118 & 128 & 140 & 153 & 170 & 189 & 205 & 222 \\ \hline
%		\end{tabular}
%	}
%\end{table}

\color{black}We implemented application with optional components, and each component has its corresponding CPU utilization and discount amount. The components are uniformly distributed on VMs.

We adopt the realistic workload trace from more than 1000 PlanetLab VMs \cite{PlanetLab} 
to create an overloaded environment \cite{AntonTPDS}.
Our experiments are simulated under one-day scheduling period and repeated for 10  different days. The brownout is invoked every 5 minutes (one time interval) if hosts are overloaded. 
The sliding window size $L_w$ in Equation (11) to estimate the number of overloaded hosts is set as 12 windows (one hour). 

\color{black}The CPU resource is measured with capacity of running instructions. Assuming that the  application workload occupies 85\% resource on a VM and the VM has 1000 million instructions per second (MIPS) computation capacity, then it represents the application constantly requires 0.85 $\times$ 1000 = 850 MI per second in the 5 minutes time interval. 
\color{black}

\begin{table}
	\centering
	\scriptsize
	\caption{Host / VM Types and Capacity}
	\resizebox{0.7\textwidth}{!}{%
		
		\begin{tabular}{|c|c|c|c|c|c|}
			
			\hline  Name & CPU & Cores  &  Memory & Bandwidth & Storage \\ 
			\hline
			\hline Host Type 1 & 1.86 GHz & 2 & 4 GB & 1 Gbit/s & 100 GB \\ 
			\hline Host Type 2 & 2.66 GHz & 2 & 4 GB & 1 Gbit/s & 100 GB  \\
			\hline
			\hline VM Type 1 & 2.5 GHz & 1 & 870 MB & 100 Mbit/s & 1 GB \\ 
			\hline VM　Type 2 & 2.0 GHz & 1 & 1740 MB & 100 Mbit/s & 1 GB  \\
			\hline VM Type 3 & 1.0 GHz & 1 & 1740 MB & 100 Mbit/s & 1 GB  \\
			\hline VM Type 4 & 0.5 GHz & 1 & 613 MB & 100 Mbit/s & 1 GB  \\				
			\hline
			
		\end{tabular}
	} 
\end{table}

We use three baseline algorithms for comparison as below:

\textbf{1) VM Placement and Consolidation  algorithm (PCO)} \cite{Beloglazov}: the algorithm has been described at the beginning of Section 4.2.

\textbf{2) Utilization-based Probabilistic VM consolidation algorithm (UBP)} \cite{Chen}: for VM initial placement, UBP adopts the same approach as PCO.
%sorting all the VMs in decreasing order based on their utilization and allocating each VM to the host that increases the least power consumption.  
For VM consolidation, UBP applies a probabilistic method \cite{Mastroianni} to select VMs from overloaded host. The probabilistic method calculates the migration probability $f_m(u)$ based on host utilization $u$ as :
$	f_m(u) = (1 - \frac{u-1}{1- T_h})^\alpha$ 
%
%
% \begin{equation}
%
%	\end{equation} 
, where $T_h$ is the upper threshold for detecting overloads and $\alpha $ is a constant to adjust probability. 

\textbf{3) Brownout algorithm with Highest Utilization and Price Ratio First Component Selection Algorithm (HUPRFCS)}\cite{Xu}: it is a brownout-based heuristic algorithm. This algorithm deactivates the application components from the one with the highest $\frac{u(App_ )}{d(App_c)}$ to the others with lower $\frac{u(App_c)}{d(App_c)}$ until the deactivated components obtain the expected utilization reduction, which is a deterministic algorithm. 
\color{black}HUPRFCS is an efficient approach to reduce energy consumption under discount amount constraints.
\color{black}

To evaluate algorithms' performance, we mainly explore two parameters: 

\textbf{1) Overloaded threshold:} it identifies the CPU utilization threshold that determines the overloaded hosts, and it is varied from 80\% to 95\% in increments of 5\%. We adopt this parameter since both \cite{Beloglazov} and \cite{Mastroianni} have shown that it influences energy consumption. 

\textbf{2) Percentage of optional utilization in an application:} it shows how much utilization in application is optional and can be deactivated. It is varied from 25\% to 100\% in increments of 25\%. \color{black} An application with 100\% optional utilization represents that the application components or microservices are self-contained and each of them is allowed to be disabled temporarily (not disabling all the components at the same time), such as a stateless online document processing application. \color{black} 
\color{black} We assume the application maximum discount is identical to the percentage of optional utilization, for example, 50\% optional utilization in an application comes along with 50\% discount amount.

\color{black} We assume that the optional components utilization $u(App_c)$ and discount $d(App_c)$ conform normal distribution $u(App_c) \AC N(\mu, \sigma^2), $ $d(App_c) \AC N(\mu, \sigma^2)$, the $\mu$ is the mean utilization of component utilization or discount, which is computed as the percentage of optional utilization (or discount amount) divided by the number of optional components. The $\sigma^2$ is the standard deviation of optional components utilization or discount. 
In our experiments, we consider both optional component utilization standard deviation and discount standard deviation are less than 0.1, which represents that the optional components are designed to have balanced utilization and discount.

\color{black}\subsection{Results}

\textbf{5.2.1 Comparison with different $\lambda$}

\begin{figure}
	\centering
	
	\includegraphics[width=0.75\linewidth]{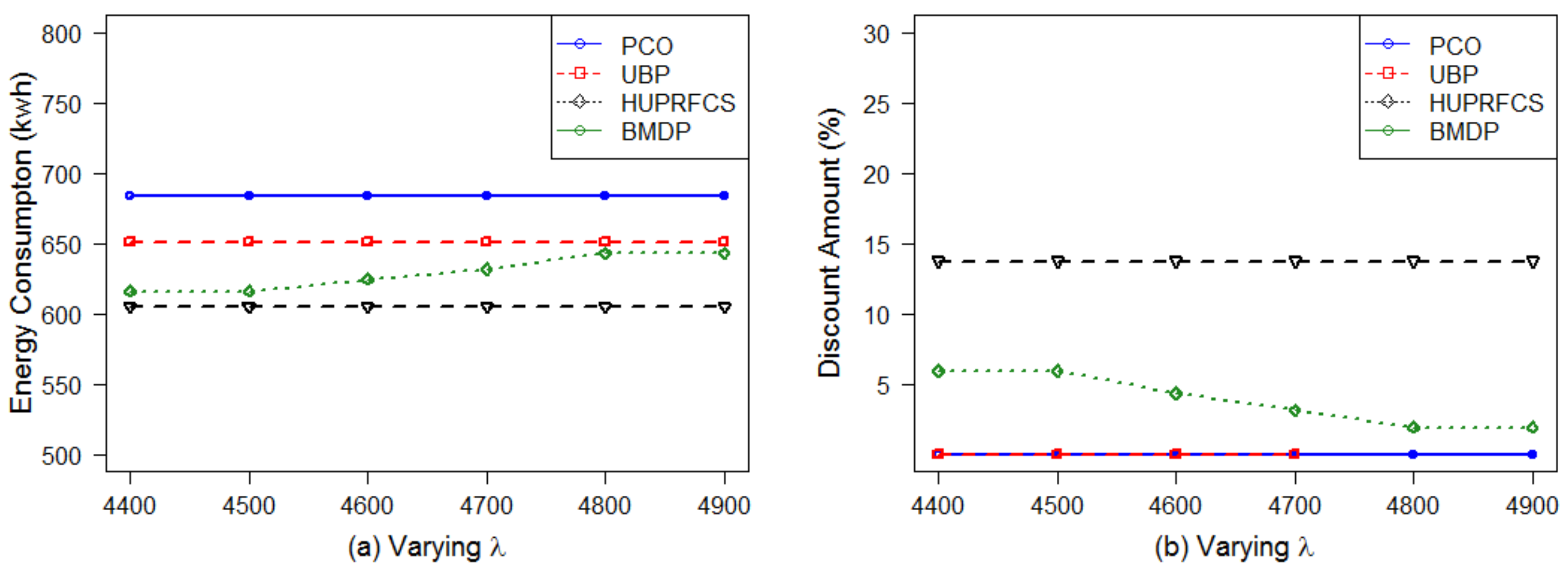}
	\caption{Comparison with different $\lambda$. 
		The parameter $\lambda$ is the weight of discount.}
\end{figure}

%\begin{table}[t]
%	\centering
%	\caption{Energy consumption and discount amount of different $\lambda$}
%	\label{my-label}
%	\resizebox{0.48\textwidth}{!}{
%		\begin{tabular}{|c|c|c|c|c|c|c|}
%			\hline
%			$\lambda$                  & 4400   & 4500   & 4600   & 4700   & 4800   & 4900   \\ \hline
%			Energy Consumption (kWh) & 615.88 & 615.88 & 624.23 & 631.50 & 642.68 & 642.68 \\ \hline
%			Discount Amount (\%)     & 5.92   & 5.92   & 4.30   & 3.16   & 1.93   & 1.93   \\ \hline
%		\end{tabular}
%	}
%\end{table}

\color{black}To investigate the impacts of different discount weights in Equation (7), we conduct a series of experiments with different $\lambda$. In these evaluations,  the hosts number and VMs number are set to 200 and 400 respectively, the overloaded threshold is set to 85\% and the percentage of optional utilization is set to 50\%. Fig. 2 indicates that energy consumption increases and discount amount decreases when $\lambda$ increases.  The reason lies in that larger $\lambda$ will guide our algorithm to find the states that offer less discount. From the results, we notice that when $\lambda$ value is less than 4500, BMDP saves more energy than UBP and PCO, and in comparison to HUPRFCS, BMDP has similar energy consumption and reduces significant discount amount.

In the following evaluations, we set $\lambda$ to a small value (i.e. $\lambda$=100) so that the energy consumption of BMDP is below two baselines (PCO and UBP) and close to HUPRFCS. Additionally, with this $\lambda$ value, the discount of BMDP is less than the discount produced by HUPRFCS. \\
\color{black}

\noindent{\textbf{5.2.2 Comparison under varied overloaded thresholds}}
\begin{figure}
\centering
	\includegraphics[width=0.75\linewidth]{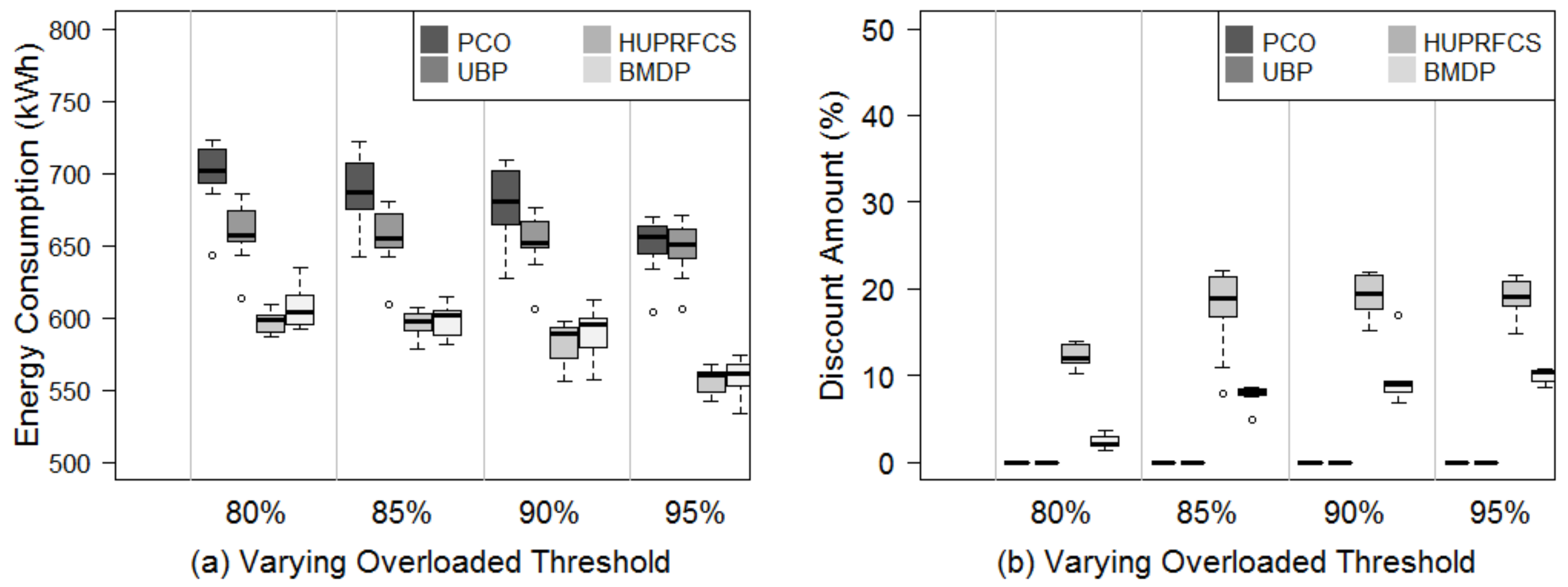}
	\caption{Varying Overloaded Threshold}
\end{figure}

\color{black}The performance evaluated under different overloaded thresholds is shown in Fig. 3. Other parameters are configured as same as in Section 5.2.1.
%, $\lambda$ in Equation (7) is set as $1.0$. 
In Fig. 3(a), we observe that the energy consumption of all the algorithms are reduced when the overloaded threshold increases, for example, PCO-80\% has 699.6 kWh with 95\% Confidence Interval (CI) (682.6, 716.6) and reduces it to 649.9 kWh with 95\% CI: (635.8, 664.1) in PCO-95\%; 
%UBP-80\% has 658.7 kWh with 95\% CI: (643.8, 673.6)  and decreases gradually to 647 kWh in UBP-95\%; 
%HUPRFCS-80\% has 598.0 kWh with 95\% CI: (592.6, 603.3) and lowers it to 557.0 kWh with 95\% CI (551.2, 562.8) in HUPRFCS-95\%; 
BMDP-80\% has 607.8 kWh with 95\% CI: (598.1, 617.4) and saves it as 558.4 kWh with 95\% CI: (549.6, 567.2) in BMDP-95\%. 
\color{black}The reason lies in that higher overloaded thresholds allow more VMs to be packed on the same host, so that more hosts are shutdown. 
When overloaded thresholds are between 80\% to 90\%, UBP reduces around 5\% energy consumption compared to PCO, while HUPRFCS and BMDP save about 14-16\% more energy consumption than PCO.  When the overloaded threshold is 95\%, PCO and UBP achieve close energy consumption, while HUPRFCS and BMDP still reduce around 16\% energy compared with them. 

As the energy consumption of HUPRFCS and BMDP are quite close, we conduct paired t-tests for HUPRFCS and BMDP as shown in Table 2. We notice that the differences between them are less than 2\%, and when the overloaded thresholds are 85\% and 95\%, the \textit{p-values} are 0.09 and 0.45 respectively, which indicates weak evidence to prove that they are different. 

Comparing the discount amount, Fig. 3(b) shows that there is no discount offered in PCO and UBP, but HUPRFCS offers 11\% to 20\% discount and BMDP reduces it to 3\% to 11\% as the trade-off due to components deactivation.  \color{black}This is because, 
%the Fig. 3(c) and Fig. 3(d), as in UBP and HUPRFCS, more utilization are reduced and more hosts are shutdown. 
based on heuristics, HUPRFCS quickly finds the components with higher utilization and discount ratio, while BMDP steps further based on MDP to optimize the component selection. %For instance, compared with HUPRFCS, BMDP deactivates the component list that costs 2\% more energy consumption but reduces 10\% discount.
\color{black}\\

\begin{table}[]
	\centering
	\caption{Paired T-Tests with 95\% CIs for Comparing Energy Consumption by HUPRFCS and BMDP under Different Overloaded Thresholds}
	\label{my-label}
	\resizebox{0.75\textwidth}{!}{%
	\begin{tabular}{|c|c|c|c|}
		\hline
		\textbf{\begin{tabular}[c]{@{}c@{}}Algorithm   1 (kWh)\end{tabular}} & \textbf{\begin{tabular}[c]{@{}c@{}}Algorithm   2 (kWh)\end{tabular}} & \textbf{\begin{tabular}[c]{@{}c@{}}Difference   (kWh)\end{tabular}} & \textbf{\textit{p-value}} \\ \hline
		\begin{tabular}[c]{@{}c@{}}HUPRFCS-80\%     (598.01)\end{tabular}    & \begin{tabular}[c]{@{}c@{}}BMDP-80\%   (607.78)\end{tabular}        & -9.77 (-15.14,  -4.39)                                                & 0.0026           \\ \hline
		\begin{tabular}[c]{@{}c@{}}HUPRFCS-85\%    (595.87)\end{tabular}     & \begin{tabular}[c]{@{}c@{}}BMDP-85\%    (599.24)\end{tabular}        & 3.37 (-0.77,  7.52)                                         & 0.099            \\ \hline
		\begin{tabular}[c]{@{}c@{}}HUPRFCS-90\%    (581.91)\end{tabular}     & \begin{tabular}[c]{@{}c@{}}BMDP-90\%    (587.97)\end{tabular}        & -6.05 (-9.41 -2.69)                                                   & 0.0027           \\ \hline
		\begin{tabular}[c]{@{}c@{}}HUPRFCS-95\%    (557.03)\end{tabular}     & \begin{tabular}[c]{@{}c@{}}BMDP-95\%    (558.41)\end{tabular}        & -1.38 (-5.36,  2.6)                                          & 0.45             \\ \hline
	\end{tabular}
}
\end{table}

\begin{figure}[t]
\centering
	\includegraphics[width=0.75\linewidth]{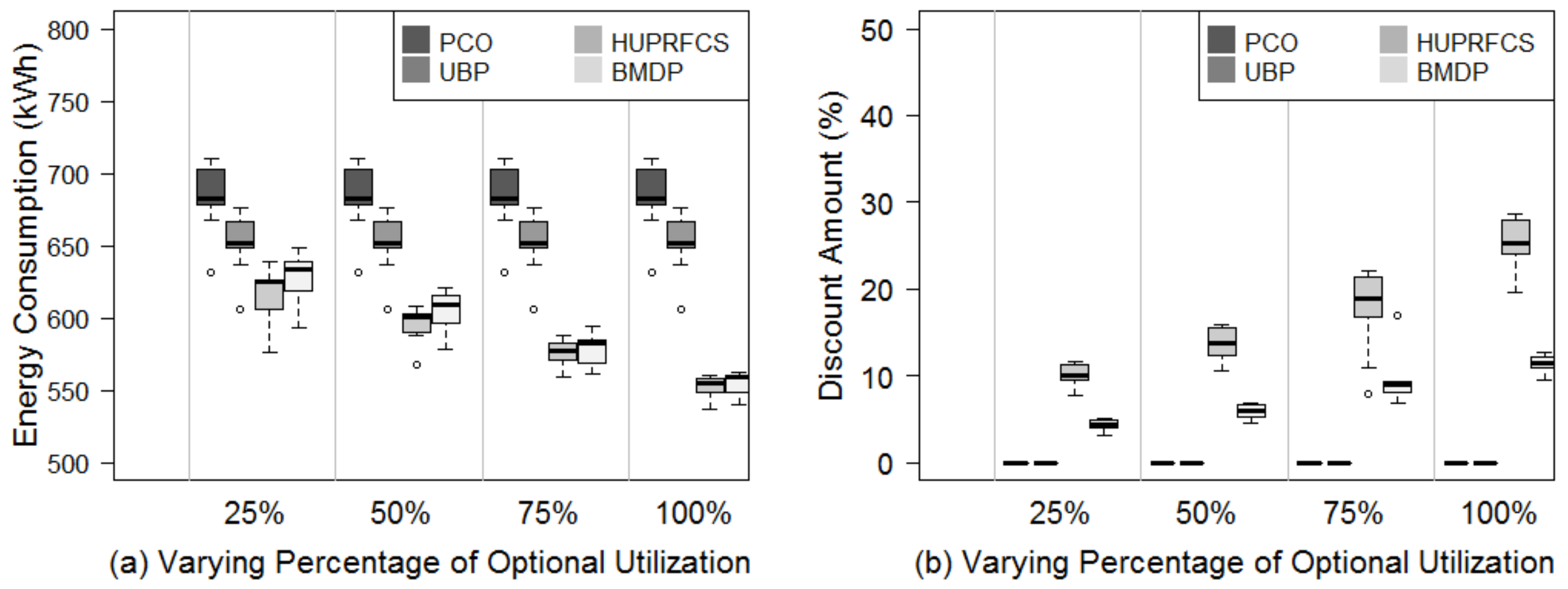}
	\caption{Varying Percentage of Optional Utilization}
\end{figure}
\noindent{\textbf{5.2.3 Comparison under Varied Percentage of Optional Utilization}}

\color{black}In Fig. 4, we compare the algorithms with different percentages of optional utilization. Other parameters are set the same as those in Section 5.2.1.
%, $\lambda$ in Equation (7) is set as $1.0$. 
As shown in Fig 4.(a), for PCO and UBP, their energy consumption are not influenced by different percentage of optional utilization.
PCO has 684 kWh with 95\% CI: (667.4, 700.6), and UBP has reduced 4.7\% to 651.9 with 95\% CI: (637.3, 666.5). 
Compared with PCO, HUPRFCS-25\% reduces 11\% energy to 605kWh with 95\% CI: (596.6, 613.4), and BMDP-25\% reduces 9\% energy to 615.9 kWh with 95\% CI: (605.9, 625.8). When the percentage of optional utilization increases, the more energy consumption is saved by HUPRFCS and BMDP. For instance, HUPRFCS-100\% and BMDP-100\% achieve around 20\% energy saving as 556.9kWh with 95\% CI: (550.9, 562.3)  and 551.6kWh with 95\% CI: (545.8, 557.4) respectively. The reason is that higher percentage of optional percentage allows more utilization to be reduced.
For the discount amount comparison in Fig. 4(b), it shows that HUPRFCS offers 10\% to 25\% discount amount as trade-offs, while BMDP only offers 3\% to 10\% discount amount. %Fig. 4(c) demonstrates that HUPRFCS provides about 3\% more disabled utilization amount than BMDP. In the comparison of the number of shutdown hosts shown in Fig. 4(d), HUPRFCS and BMDP shut down about 18-20 more host than PCO and 3-5 more hosts than UBP. 

Because the energy consumption of HUPRFCS and BMDP are quite close, we conduct the paired t-test for HUPRFCS and BMDP as illustrated in Table 3. When the percentage of optional utilization are 75\% and 100\%, the \textit{p-values} are 0.099 and 0.057, which indicates weak evidence to prove that they are different. And with other percentage of optional utilization, the energy consumption differences are less than 2\%. 

%As a result, BMDP saves the approximate energy consumption as HUPRFCS and reduces the discount given to users, which shows that BMDP achieves better trade-offs between energy consumption and discount.  \color{black}

\begin{table}[]
	\centering
	\caption{Paired T-Tests with 95\% CIs for Comparing Energy Consumption by HUPRFCS and BMDP under Different Percentage of Optional Utilization}
	\label{my-label}
	\resizebox{0.75\textwidth}{!}{%
	\begin{tabular}{|c|c|c|c|}
		\hline
		\textbf{\begin{tabular}[c]{@{}c@{}}Algorithm   1 (kWh)\end{tabular}} & \textbf{\begin{tabular}[c]{@{}c@{}}Algorithm   2 (kWh)\end{tabular}} & \textbf{\begin{tabular}[c]{@{}c@{}}Difference   (kWh)\end{tabular}} & \textbf{\textit{p-value}} \\ \hline
		\begin{tabular}[c]{@{}c@{}}HUPRFCS-25\%     (617.57)\end{tabular} & \begin{tabular}[c]{@{}c@{}}BMDP-25\%    (628.10)\end{tabular}        & -10.52 (-12.52,  -8.52)                                               & 0.00082          \\ \hline
		\begin{tabular}[c]{@{}c@{}}HUPRFCS-50\%    (595.0)\end{tabular}   & \begin{tabular}[c]{@{}c@{}}BMDP-50\%    (605.88)\end{tabular}        & -10.88 (-15.26, -6.5)                                                 & 0.00032          \\ \hline
		\begin{tabular}[c]{@{}c@{}}HUPRFCS-75\%    (575.87)\end{tabular}  & \begin{tabular}[c]{@{}c@{}}BMDP-75\%    (579.24)\end{tabular}        & -3.37 (-7.52 -0.78)                                                   & 0.099            \\ \hline
		\begin{tabular}[c]{@{}c@{}}HUPRFCS-100\%    (551.56)\end{tabular} & \begin{tabular}[c]{@{}c@{}}BMDP-100\%    (556.59)\end{tabular}       & -3.12 (-5.08,  -1.16)                                                 & 0.0057           \\ \hline
	\end{tabular}
}
\end{table}

%\begin{figure}[!ht]
%
%\centering
%	\includegraphics[width=0.5\linewidth]{ConvergenceTime}
%	\caption{Convergence Time Comparison with Different Component Numbers.		
%	$M$ is the number of PMs, $N$ is the number of VMs. $C$ is the maximum number of components on VMs.}
%\end{figure}
%Table 1 shows the convergence time of BMDP under different number of PMs, VMs and maximum components on VMs. The results show that our algorithm executes in an acceptable time.
%
%\begin{table}[H]
%	\centering
%	\caption{Scalability of BMDP}
%	\label{my-label}
%	\begin{tabular}{|c|c|c|c|c|c|c|c|}
%		\hline
%		\backslashbox{$|C|$}{$|M|\cdot|N|$} & 20,000 & 45,000 & 80,000 & 125,000 & 180,000 & 245,000 & 320,000\\ \hline
%10 & 29s            & 65s         & 92s         & 125s        & 139s         & 169s         & 194s         \\ \hline
%20 & 40s            & 79s         & 116s        & 146s        & 180s         & 215s         & 262s         \\ \hline
%30 & 85s            & 126s        & 190s        & 241s        & 291s         & 347s         & 417s         \\ \hline
%
%	\end{tabular}
%\end{table}

\color{black}

\section{Conclusions and Future Work}\label{sec:Conclusion}
\color{black}Brownout has been proven to be effective to solve the overloaded situation in cloud data centers. Additionally, brownout can also be applied to reduce energy consumption. 
\color{black}In this paper, we introduced the brownout system model by deactivating optional components in applications or microservices temporarily. In the model, the brownout controller can deactivate the optional components or microservices to deal with overloads and reduce data center energy consumption while offering discount to users. We also propose an algorithm based on brownout and approximate Markov Decision Process namely BMDP, to find the components should be deactivated. The simulations based on real trace showed that BMDP reduces 20\% energy consumption than non-brownout baselines and saves discount amount than brownout baseline. As future work, we plan to implement a brownout prototype based on Docker Swarm.
\\

%\section*{Acknowledgments}\label{sec:Acknowledgments}
%
\noindent{\textbf{Acknowledgments}.\addcontentsline{toc}{section}{Acknowledgment}
This work is supported by China Scholarship Council, Australia Research Council Future Fellowship and Discovery Project Grants. 
We thank Chenhao Qu, Adel Nadjaran Toosi 
and Satish Narayana Srirama 
for their valuable suggestions.}

\bibliographystyle{splncs03}

\bibliography{mdp}

%\begin{thebibliography}{1}
%
%\bibitem{Einstein}
%A. Einstein, On the movement of small particles suspended in stationary liquids required by the molecular-kinetic theory of heat, Annalen der Physik 17, pp. 549-560, 1905.
%
%\end{thebibliography}

\end{document}